\documentclass[Afour,sageh,times]{sagej}

\usepackage{moreverb,url}

\usepackage[colorlinks,bookmarksopen,bookmarksnumbered,citecolor=red,urlcolor=red]{hyperref}

\usepackage{listings}
\graphicspath{{figures/}}

\usepackage{multirow}
\usepackage{xcolor}

\lstdefinelanguage{MyC}
{
  keywords=[1]{inline, restrict, static, double, const, int, for, void},
  keywordstyle=[1]\bfseries\color{green!40!black},
  keywords=[2]{__restrict__, __attribute__},
  keywordstyle=[2]\bfseries\color{blue!40!black},
  morecomment=[l][\itshape\color{purple!40!black}]{//},
  morecomment=[l][\bfseries\color{blue!40!black}]{\#},
}

\newcommand\BibTeX{{\rmfamily B\kern-.05em \textsc{i\kern-.025em b}\kern-.08em
T\kern-.1667em\lower.7ex\hbox{E}\kern-.125emX}}

\def\volumeyear{2019}

\begin{document}

\runninghead{Sun et al.}

\title{A study of vectorization for matrix-free finite element methods}

\author{Tianjiao Sun\affilnum{1}, Lawrence Mitchell\affilnum{2}, Kaushik Kulkarni\affilnum{3}, Andreas Kl\"{o}ckner\affilnum{3}, David A. Ham\affilnum{4}, and Paul H. J. Kelly\affilnum{1}}

\affiliation{\affilnum{1}Department of Computer Science, Imperial College London, UK\\
\affilnum{2}Department of Computer Science, Durham University, UK\\
\affilnum{3}Department of Computer Science, University of Illinois at Urbana-Champaign, USA\\
\affilnum{4}Department of Mathematics, Imperial College London, UK}

\corrauth{Tianjiao Sun, Imperial College London,
Department of Computing,
London,
SW7~2AZ, UK.}

\email{me@tjsun.info}

\begin{abstract}
Vectorization is increasingly important to achieve high performance on modern hardware with SIMD instructions. Assembly of matrices and vectors in the finite element method, which is characterized by iterating a local assembly kernel over unstructured meshes, poses difficulties to effective vectorization. Maintaining a user-friendly high-level interface with a suitable degree of abstraction while generating efficient, vectorized code for the finite element method is a challenge for numerical software systems and libraries. In this work, we study cross-element vectorization in the finite element framework Firedrake via code transformation and demonstrate the efficacy of such an approach by evaluating a wide range of matrix-free operators spanning different polynomial degrees and discretizations on two recent CPUs using three mainstream compilers. Our experiments show that our approaches for cross-element vectorization achieve 30\% of theoretical peak performance for many examples of practical significance, and exceed 50\% for cases with high arithmetic intensities, with consistent speed-up over (intra-element) vectorization restricted to the local assembly kernels.
\end{abstract}


\maketitle

\section{Introduction}

The realization of efficient solution procedures for partial differential equations (PDEs) using finite element methods on modern computer systems requires the combination of diverse skills across mathematics, programming languages and high-performance computing. Automated code generation is one of the promising approaches to manage this complexity. It has been increasingly adopted in software systems and libraries. Recent successful examples include FEniCS~\citep{logg2012automated}, Firedrake~\citep{rathgeber2015firedrake} and FreeFem++~\citep{hecht2012new}. These software packages provide users with high-level interfaces for high productivity while relying on optimizations and transformations in the code generation pipeline to generate efficient low-level code. The challenge, as in all compilers, is to use appropriate abstraction layers that enable optimizations to be applied that achieve high performance on a broad set of programs and machines.

One particular challenge for generating high-performance code on modern hardware is vectorization. Modern CPUs increasingly rely on SIMD instructions to achieve higher throughput and better energy efficiency. Finite element computation requires the assembly of vectors and matrices which represent differential forms on discretized function spaces. This process consists of applying a local function, often called an \textit{element kernel}, to each mesh entity, and incrementing the global data structure with the local contribution. Typical local assembly kernels suffer from issues that can preclude effective vectorization. These issues include complicated loop structures, poor data access patterns, and short loop trip counts that are not multiples of the vector width. As we show in this paper, general purpose compilers perform poorly in generating efficient, vectorized code for such kernels. Padding and data layout transformations are required to enable the vectorization of the element kernels~\citep{luporini2015cross}, but the effectiveness of such approaches is not consistent across different examples. Since padding may also result in larger overheads for wider vector architectures, new strategies are needed as vector width increases for the new generation of hardware.

Matrix-free methods avoid building large sparse matrices in applications of the finite element method and thus trade computation for storage. They have become popular for use on modern hardware due to their higher arithmetic intensity (defined as the number of floating-point operations per byte of data transfer). Vectorization is particularly important for computationally intensive high order methods, for which matrix-free methods are often applied. Previous work on improving vectorization of matrix-free operator application, or equivalently, residual evaluation, mostly focuses on exposing library interfaces to the users. \citet{kronbichler2017fast} first perform a change of basis from nodal points to quadrature points, and provide overloaded SIMD types for users to write a quadrature-point-wise expression for residual evaluation. However, since the transformation is done manually, new operators require manual reimplementation. \citet{knepley2013finite} also transpose to quadrature-point basis but target GPUs instead. Both works vectorize by grouping elements into batches, either to match the SIMD vector length in CPUs or the shared memory capacity on GPUs. In contrast, \citet{muthing2017high} apply an intra-kernel vectorization strategy and exploit the fact that, in 3D, evaluating both a scalar field and its three derivatives fills the four lanes of an AVX2 vector register (assuming the computation is in double precision). More recently, \citet{kempf2018automatic} target high order Discontinuous Galerkin (DG) methods on hexahedral meshes using automated code generation to search for vectorization strategies, while taking advantage of the specific memory layout of the data.

In this work, we present a generic and portable solution based on cross-element vectorization. Our vectorization strategy, implemented in Firedrake, is similar to that of~\citet{kronbichler2017fast} but is fully automated through code generation like that of~\citet{kempf2018automatic}. We extend the scope of code generation in Firedrake to incorporate the outer iteration over mesh entities and leverage Loopy~\citep{klockner2014loo}, a loop code generator based loosely on the polyhedral model, to systematically apply a sequence of transformations which promote vectorization by grouping mesh entities into batches so that each SIMD lane operates on one entity independently. This automated code generation mechanism enables us to explore the effectiveness of our techniques on operators spanning a wide range of complexity and systematically evaluate our methodology. Compared with an intra-kernel vectorization strategy, this approach is conceptually well-defined, more portable, and produces more predictable performance. Our experimental evaluation demonstrates that the approach consistently achieves a high fraction of hardware peak performance while being fully transparent to end users.

The contributions of this work are as follows:
\begin{itemize}
\item We present the design of a code transformation pipeline that permits the generation of high-performance, vectorized code on a broad class of finite element models.
\item We demonstrate the implementability of the proposed pipeline by realizing it in the Firedrake finite element framework.
\item We provide a thorough evaluation of our code generation strategy and demonstrate that it achieves a substantial fraction of theoretical peak performance across a broad range of test cases and popular C compilers.
\end{itemize}

The rest of this paper is arranged as follows. After reviewing the preliminaries of code generation for the finite element method in Section~\ref{sec:preliminaries}, we describe our implementation of cross-element vectorization in Firedrake in Section~\ref{sec:vectorization}. In Section~\ref{sec:evaluation}, we demonstrate the effectiveness of our approach with experimental results. Finally, we review our contributions and identify future research priorities in Section~\ref{sec:conclusion}.

\section{Preliminaries}%
\label{sec:preliminaries}

The computation of multilinear forms using the basis functions spanning the discretized function spaces is called \textit{finite element assembly}. When applying the matrix-free methods, one only needs to assemble linear forms, or residual forms, because matrix-vector products are essentially the assembly of linear forms which represent the \textit{actions} of bilinear forms. Optimizing linear form assembly is therefore crucial for improving the performance of matrix-free methods. In Firedrake, one can invoke the matrix-free approach without changing the high-level problem formulation by setting solver options as detailed by~\citet{kirby2018solver}.

The general structure of a linear form $L$ is
\begin{equation}
L(c_1,c_2,\ldots,c_k;v):\hat{V}_1\times\hat{V}_2\times\ldots\hat{V}_k\times V\rightarrow \mathrm{R},
\end{equation}
where $c_i\in\hat{V_i},i=1\ldots k$, are arbitrary coefficient functions, and $v\in V$ is the test function. $L$ is linear with respect to $v$, but possibly nonlinear with respect to the coefficient functions.

Let ${\{\phi_i\}_{i=1}^N}$ be the set of basis functions spanning $V$. Define $v_i=L(c_1,\ldots,c_k; \phi_i)\in \mathrm{R}$, then the assembly of $L$ constitutes the computation of the vector $\mathbf{v}=(v_i, \ldots,v_n)$. In Firedrake, this is treated as a two-step process: \textit{local} assembly and \textit{global} assembly. The rest of this section highlights the computational properties of these two steps with an example, and in doing so, introduces the components and concepts in Firedrake that are relevant to the implementation of cross-element vectorization.

\subsection{Local assembly}

Local assembly of linear forms is the evaluation of the integrals as defined by the weak form of the differential equation on each entity (cell or facet) of the mesh. In Firedrake, the users define the problem in \textit{Unified Form Language} (UFL)~\citep{alnaes2014unified} which captures the weak form and the function space discretization. Then the \textit{Two-Stage Form Compiler} (TSFC)~\citep{homolya2018tsfc} takes this high-level, mathematical description and generates efficient C code. The intermediate representation of TSFC is a tensor algebra language called \textit{GEM}, which supports various optimizations on the tensor operations. As an example, consider the linear form of the weak form of the positive-definite Helmholtz operator:
\begin{equation}
L(u;v) = \int_{\Omega}\nabla u\cdot \nabla v + uv\ \mathrm{d}x,
\label{eq:helmholtz-bilinear-form}
\end{equation}

\begin{lstlisting}[language=Python,
morekeywords={UnitSquareMesh, FunctionSpace, TestFunction, Function, dot, grad, dx, assemble},
basicstyle=\scriptsize\ttfamily,
keywordstyle=\bfseries\color{green!40!black},
commentstyle=\itshape\color{purple!40!black},
firstline=1,
lastline=6,
firstnumber=1,
frame=single,
showspaces=false,
showstringspaces=false,
numbers=right,
breaklines=true,
caption=Assembling the linear form of the Helmholtz operator in UFL.,
label=lst:helmholtz_ufl]
mesh = UnitSquareMesh(10, 10)
V = FunctionSpace(mesh, "Lagrange", 2)
v = TestFunction(V)
u = Function(V)
L = (dot(grad(u), grad(v)) + u*v) * dx
result = assemble(L)
\end{lstlisting}

\begin{lstlisting}[language=MyC,
basicstyle=\scriptsize\ttfamily,
float=*htbp,
firstnumber=1,
frame=single,
showspaces=false,
showstringspaces=false,
numbers=left,
breaklines=true,
caption=Local assembly kernel for the Helmholtz operator of Listing~\ref{lst:helmholtz_ufl} in C generated by TSFC.,
label=lst:helmholtz_local]
static inline void helmholtz(double *__restrict__ A, double const *__restrict__ coords,
                             double const *__restrict__ w_0)
{
  double const t13[6] = { ... };
  double const t14[6 * 6] = { ... };
  double const t15[6 * 6] = { ... };
  double const t16[6 * 6] = { ... };
  double t0 = -1.0 * coords[0];
  double t1 = t0 + coords[2];
  double t2 = -1.0 * coords[1];
  double t3 = t2 + coords[5];
  double t4 = t0 + coords[4];
  double t5 = t2 + coords[3];
  double t6 = t1 * t3 + -1.0 * t4 * t5;
  double t7 = 1.0 / t6;
  double t8 = t1 * t7;
  double t9 = t7 * -1.0 * t4;
  double t10 = t7 * -1.0 * t5;
  double t11 = t3 * t7;
  double t12 = fabs(t6);
  for (int ip = 0; ip <= 5; ++ip) {
    double t17 = 0.0;
    double t18 = 0.0;
    double t19 = 0.0;
    for (int i = 0; i <= 5; ++i) {
      t17 = t17 + t16[6 * ip + i] * w_0[i];
      t18 = t18 + t15[6 * ip + i] * w_0[i];
      t19 = t19 + t14[6 * ip + i] * w_0[i];
    }
    double t20 = t13[ip] * t12;
    double t21 = t11 * t19 + t10 * t18;
    double t22 = t9 * t19 + t8 * t18;
    double t23 = t20 * (t21 * t10 + t22 * t8);
    double t24 = t20 * (t21 * t11 + t22 * t9);
    double t25 = t20 * t17;
    for (int j = 0; j <= 5; ++j)
      A[j] = A[j] + t15[6 * ip + j] * t23 + t16[6 * ip + j] * t25 + t14[6 * ip + j] * t24;
  }
}
\end{lstlisting}

Listing~\ref{lst:helmholtz_ufl} shows the UFL syntax to assemble the linear form $L$ as the vector \texttt{result}, on a $10\times10$ triangulation of a unit square. We choose to use the second-order Lagrange element, commonly known as the \texttt{P2} element, as our approximation space. Listing~\ref{lst:helmholtz_local} shows a C representation of this kernel generated by TSFC\endnote{This TSFC-generated kernel is reformatted slightly for consistency with the PyOP2 generated kernels. The kernel function names generated by TSFC and PyOP2 are long and complicated due to name mangling in Firedrake. They are shortened to operator names such as \texttt{helmholtz} in the listings for readability.}. We note the following key features of this element kernel:
\begin{itemize}
\item The kernel takes three array arguments in this case: \texttt{coords} holds the coordinates of the current triangle, \texttt{w\_0} holds ${u_i}$, the coefficients of $u$, and \texttt{A} stores the result.
\item The first part of the kernel (lines 8--20) computes the inverse and the determinant of the Jacobian for the coordinate transformation from the reference element to the current element. This is required for \textit{pulling back} the differential forms to the reference element. The Jacobian is constant for each triangle because the coordinate transformation is affine in this case. In the general case, the Jacobian is computed at every quadrature point.
\item The constant arrays \texttt{t13}, \texttt{t14}, \texttt{t15}, \texttt{t16} are the same for all elements. \texttt{t14} represents the tabulation of basis functions at quadrature points, \texttt{t15} and \texttt{t16} represent derivatives of basis functions at quadrature points across each spatial dimension, \texttt{t13} represents the quadrature weights.
\item The \texttt{ip} loop iterates over the quadrature points, evaluating the integrand in~\eqref{eq:helmholtz-bilinear-form} and summing to approximate the integral. The \texttt{i} and \texttt{j} loops iterate over the degrees of freedom performing a change of basis to values at quadrature points and then back to degrees of freedom when accumulating into the output array \texttt{A}. The extents of these loops depend on the integrals performed and the choice of function spaces respectively.
\item TSFC performs a sequence of optimization passes by rewriting the tensor operations following mathematical rules before generating the loop nests (see \citet{homolya2017exposing} for details). This is more powerful than the C compiler's \textit{loop-invariant code motion} optimization because, firstly, TSFC operates on the symbolic tensor expressions to explore different refactoring and reordering strategies that expose invariant sub-expressions, and secondly, non-scalar sub-expressions can also be extracted into temporary arrays to eliminate redundant computation in the loop nests. As a side effect of these transformations, the loop nests in the kernels are no longer perfectly nested, thus limiting the effectiveness of vectorization if it is only applied to the innermost loops.
\item After the optimization and scheduling stage, the translation of tensor algebra to C in TSFC is a straightforward rewrite of tensor operations to loop nests. This process results in certain artefacts that are shown in Listing~\ref{lst:helmholtz_local}. For example, since there is no specific representation for subtractions in TSFC, negations are emitted as multiplication by \texttt{-1}. This can result in generated C code that is not as idiomatic as if written by hand. We rely on the modern C compilers to optimize such artefacts away, since readability is a secondary concern for the generated code.

\end{itemize}

\subsection{Global assembly}

During global assembly, the local contribution from each mesh entity, computed by the element kernel, is accumulated into the global data structure. In Firedrake, PyOP2~\citep{Rathgeber2012} is responsible for representing and realizing the iteration over mesh entities, marshalling data in and out of the element kernels. The computation is organized as PyOP2 parallel loops, or \textit{parloops}. A parloop specifies a computational kernel, a set of mesh entities to which the kernel is applied, and all data required for the kernel. The data objects can be directly defined on the mesh entities, or indirectly accessed through maps from the mesh entities. For instance, the signature for the global assembly of the Helmholtz operator is:
\begin{lstlisting}[language=python,basicstyle=\scriptsize\ttfamily,frame=none,morekeywords={parloop}]
parloop(helmholtz, cells, L(cell2dof, INC), 
        coords(cell2vert, R), u(cell2dof, R)).
\end{lstlisting}
Here \texttt{helmholtz} is the element kernel as shown in Listing~\ref{lst:helmholtz_local}, generated by TSFC; \texttt{cells} is the set of all triangles in the mesh; \texttt{L}, \texttt{coords}, and \texttt{u} are the global data objects that are needed to create the arguments for the element kernel, where \texttt{L} holds the result vector, \texttt{coords} holds the coordinates of the vertices of the triangles which are needed for computing the Jacobian, and \texttt{u} holds the vector representation of function $u$ (as weights of basis functions). These global data objects correspond to the kernel arguments \texttt{A}, \texttt{coords} and \texttt{w\_0} respectively. The maps \texttt{cell2dof} and \texttt{cell2vert} provide indirection from mesh entities to the global data objects, and each data argument is annotated with an access descriptor (\texttt{R} for read-only, \texttt{INC} for increment access). In this example, the \texttt{L} and \texttt{u} arguments share the same map (since they are both defined on the same quadratic Lagrange space) while the \texttt{coords}, being linear, use a different map.

\begin{lstlisting}[language=MyC,
basicstyle=\scriptsize\ttfamily,
float=*!htbp,
firstnumber=1,
frame=single,
showspaces=false,
showstringspaces=false,
numbers=left,
breaklines=true,
caption=Global assembly code for action of the Helmholtz operator in C generated by PyOP2.,
label=lst:helmholtz_global]
static inline void helmholtz(double *__restrict__ A, double const *__restrict__ coords,
                             double const *__restrict__ w_0)
{
  // ... element kernel as defined previously ... //
}

void wrap_helmholtz(int const start, int const end, double *__restrict__ dat0, double const *__restrict__ dat1,
                    double const *__restrict__ dat2, int const *__restrict__ map0, int const *__restrict__ map1)
{
  double t2[6];
  double t3[3 * 2];
  double t4[6];

  for (int n = start; n <= -1 + end; ++n) {
    for (int i6 = 0; i6 <= 5; ++i6)
      t4[i6] = dat2[map0[6 * n + i6]];

    for (int i2 = 0; i2 <= 2; ++i2)
      for (int i3 = 0; i3 <= 1; ++i3)
        t3[2 * i2 + i3] = dat1[2 * map1[3 * n + i2] + i3];

    for (int i1 = 0; i1 <= 5; ++i1)
      t2[i1] = 0.0;

    helmholtz(t2, t3, t4);

    for (int i15 = 0; i15 <= 5; ++i15)
      dat0[map0[6 * n + i15]] += t2[i15];
  }
}
\end{lstlisting}
Listing~\ref{lst:helmholtz_global} shows the C code generated by PyOP2 for the above example. The code is then JIT-compiled when the result is needed in Firedrake. In the context of vectorization, this approach, with the inlined element kernel, forms the baseline in our experimental evaluation. We note the following key features of the global assembly kernel:
\begin{itemize}
\item The outer loop is over mesh entities.
\item For each entity, the computation can be divided into three parts: gathering the input data from global into local data (\texttt{t3} and \texttt{t4} in this case, which correspond to kernel arguments \texttt{coords} and \texttt{w\_0}), calling the local assembly kernel, scattering the output data (\texttt{t2}) to the global data structure.
\item The gathering and scattering of data make use of indirect addressing via base pointers (\texttt{dat}s) and indices (\texttt{map}s).
\item Different mesh entities might share the same degrees of freedom: parallelization of the scattering loop on line 27 must be aware of the potential for data races.
\item Global assembly interacts with local assembly via a function call (line 25). While the C compiler can inline this call, it creates an artificial boundary to using loop optimization techniques that operate at the source code level. Additionally, even after inlining, outer loop vectorization over mesh entities requires that the C compiler vectorize through data-dependent array accesses. This is the software engineering challenge that has previously limited vectorization to a single local assembly kernel in Firedrake.
\end{itemize}

\section{Vectorization}%
\label{sec:vectorization}

As one would expect, the loop nests and loop trip counts vary considerably for different integrals, meshes and function spaces that users might choose. This complexity is one of the challenges that our system specifically, and Firedrake more generally, must face in order to deliver predictable performance on modern CPUs, which have increasingly rich SIMD instruction sets.

In the prior approach to vectorization in our framework, the local assembly kernels generated by TSFC were further transformed to facilitate vectorization, as described in~\citet{luporini2015cross}. The arrays are padded so that the trip counts of the innermost loops match multiples of the length of SIMD units. However, padding becomes less effective for low polynomial degrees on wide SIMD units. For instance, AVX512 instructions act on 8 double-precision floats, but the loops for degree 1 polynomials on triangles only have trip counts of 3. Moreover, loop-invariant code motion is very effective in reducing the number of floating-point operations, but hoisted instructions are not easily vectorized as they are no longer in the innermost loops. This effect is more pronounced on tensor-product elements where TSFC is able to apply \textit{sum factorization}~\citep{homolya2017exposing} to achieve better algorithmic complexity.

\subsection{Cross-element vectorization and Loopy}%
\label{subsection:openmp_pragma}

Another strategy is to vectorize across several elements in the outer loop over the mesh entities, as proposed previously by~\citet{kronbichler2017fast}. This approach computes the contributions from several mesh entities using SIMD instructions, where each SIMD lane handles one entity. This is always possible regardless of the complexity of the local element kernel because the computation on each entity is independent and identical. One potential downside is the increase in register and cache pressure as the working set is larger.

For a compiler, the difficulty in performing cross-element vectorization (or, more generally, outer-loop vectorization) is to automate a sequence of loop transformations and necessary data layout transformations robustly. This is further complicated by the indirect memory access in data gathering and scattering, and the need to unroll and interchange loops across the indirections, which requires significantly more semantic knowledge than that available to the C compiler.

Loopy~\citep{klockner2014loo} is a loop generator embedded in Python which targets both CPUs and GPUs. Loopy provides abstractions based on integer sets for loop-based computations and enables powerful transformations based on the polyhedral model~\citep{verdoolaege2010isl}. Loop-based computations in Loopy are represented as \textit{Loopy kernels}. A Loopy kernel is a subprogram consisting of a loop domain and a partially-ordered list of scalar assignments acting on multi-dimensional arrays. The loop domain is specified as the set of integral points in the convex intersection of quasi-affine constraints, represented using the Integer Set Library~\citep{verdoolaege2010isl}. Loopy supports code generation for different environments from the same kernel by choosing different \emph{targets}. 

\begin{figure*}[htbp!]
  \includegraphics[width=\textwidth,keepaspectratio]{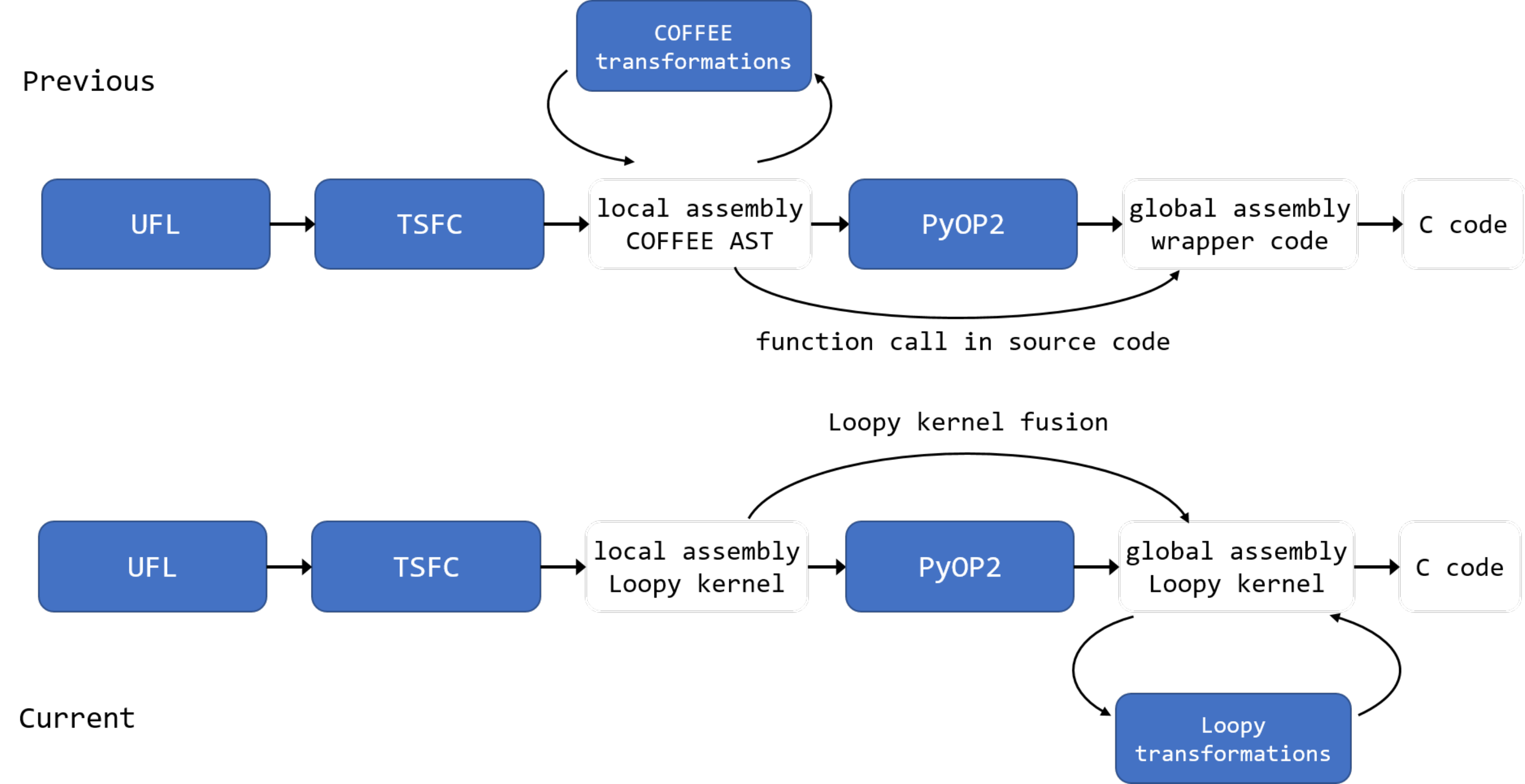}
  \caption{Integration of Loopy in Firedrake for global assembly code generation.}%
  \label{fig:loopy_integration}
\end{figure*}

To integrate with Loopy, the code generation mechanisms in Firedrake were modified as illustrated in Figure~\ref{fig:loopy_integration}.
Instead of generating source code directly, TSFC and PyOP2 are modified to generate Loopy kernels. We have augmented the Loopy internal representation with the ability to support a generalized notion of kernel fusion through the nested composition of kernels, specifically through subprograms and inlining. This allows PyOP2 to inline the element kernel such that the global assembly Loopy kernel encapsulates the complete computation of global assembly. This holistic view of the overall computation enables robust loop transformations for vectorization across the boundary between global and local assembly. To facilitate SIMD instruction generation, we also introduced a new OpenMP target to Loopy which extends its existing C-language target to support OpenMP SIMD directives~\citep[\S2.9.3]{openmp18}.

Listing~\ref{lst:helmholtz_loopy} shows an abridged version of the global assembly Loopy kernel for the Helmholtz operator, with the element kernel fused.
\begin{lstlisting}[language=c,
morekeywords={KERNEL, ARGUMENTS, DOMAINS, INAME_IMPLEMENTATION_TAGS, TEMPORARIES, INSTRUCTIONS},
deletekeywords={for},
basicstyle=\scriptsize\ttfamily,
keywordstyle=\bfseries\color{green!40!black},
commentstyle=\itshape\color{purple!40!black},
float=htbp,
firstnumber=1,
frame=single,
showspaces=false,
showstringspaces=false,
numbers=right,
breaklines=true,
caption=Global assembly Loopy kernel of the Helmholtz operator.,
label=lst:helmholtz_loopy]
KERNEL: helmholtz
---------------------------------------------------------
ARGUMENTS:
start: type: int32
end: type: int32
dat0: type: float64, shape: (None)
// ... More arguments ... //
---------------------------------------------------------
DOMAINS:
[end, start] -> { [n] : start <= n < end }
{ [i6] : 0 <= i6 <= 5 }
// ... More domains ... //
---------------------------------------------------------
INAME_IMPLEMENTATION_TAGS:
None
---------------------------------------------------------
TEMPORARIES:
t4: type: float64, shape: (6), dim_tags: (stride:1)
// ... More temporaries ... //
---------------------------------------------------------
INSTRUCTIONS:
for n
  for i6   
    t4[i6] = dat2[map0[n, i6]]
  // ... More instructions ... //
  for i15
    dat0[map0[n, i15]] += t0[0, i15]
end n
\end{lstlisting}
We highlight the following key features of Loopy kernels:
\begin{itemize}
\item Loop indices, such as \texttt{n} and \texttt{i1}, are called \emph{inames} in Loopy, which define the iteration space. The bounds of the loops are specified by the affine constraints in \emph{domains}.
\item Loop transformations operate on kernels by rewriting the loop domain and the statements making up the kernel. In addition, each iname carries a set of \emph{tags} governing its realization in generated code, perhaps as a sequential loop, as a vector lane index, or through unrolling.
\item Multi-dimensional arrays occur as \emph{arguments} and \emph{temporaries}. The memory layout of the data can be specified by assigning \emph{tags} to the array dimensions.
\item Dependencies between statements specify their partial order. Statement scheduling can also be controlled by assigning priorities to statements and inames.
\end{itemize}
For example, to achieve cross-element vectorization (by batching four elements into one SIMD vector in this example) we invoke the following sequence of Loopy transformations on the global assembly Loopy kernel, exploiting the domain knowledge of finite element assembly:
\begin{enumerate}
\item Split the outer loop \texttt{n} over mesh entities into \texttt{n\_outer} and \texttt{n\_simd}, with \texttt{n\_simd} having a trip count of four. The objective is to generate SIMD instructions for the \texttt{n\_simd} loops, such that each vector lane computes one iteration of the \texttt{n\_simd} loops.
\item Assign the tag \texttt{SIMD} to the new iname \texttt{n\_simd}. This tag informs Loopy to force the \texttt{n\_simd} loop to be innermost, privatizing data by vector-expansion if necessary.
\end{enumerate}
We show the change to the Loopy kernel after these transformations in Listing~\ref{lst:helmholtz_loopy_vector}. In particular, the vector-expansion of the temporary \texttt{t4}, and the splitting (and subsequent modification of the loop domain) of the \texttt{n} iname. 
\begin{lstlisting}[language=c,
morekeywords={KERNEL, ARGUMENTS, DOMAINS, INAME_IMPLEMENTATION_TAGS, TEMPORARIES, INSTRUCTIONS},
deletekeywords={for},
basicstyle=\scriptsize\ttfamily,
keywordstyle=\bfseries\color{green!40!black},
commentstyle=\itshape\color{purple!40!black},
float=htbp,
firstnumber=1,
frame=single,
showspaces=false,
showstringspaces=false,
numbers=right,
breaklines=true,
caption=Changes to global assembly Loopy kernel of the Helmholtz operator after cross-element vectorization,
label=lst:helmholtz_loopy_vector]
KERNEL: helmholtz_simd
---------------------------------------------------------
ARGUMENTS:
start: type: int32
end: type: int32
dat0: type: float64, shape: (None)
// ... More arguments ... //
---------------------------------------------------------
DOMAINS:
[end, start] -> { [n_outer, n_simd] :
  n_simd >= start - 4n_outer
  and 0 <= n_simd <= 3
  and n_simd < end - 4n_outer }
// ... More domains ... //
---------------------------------------------------------
INAME_IMPLEMENTATION_TAGS:
n_simd: SIMD
---------------------------------------------------------
TEMPORARIES:
t4: type: float64, shape: (6, 4),
          dim_tags: (stride:4, stride:1)
// ... More temporaries ... //
---------------------------------------------------------
INSTRUCTIONS:
for n_outer, n_simd
  for i6
    t4[i6, n_simd] = dat2[map0[n_outer*4 + n_simd, i6]]
  // ... More instructions ... //
  for i15
    dat0[map0[n_outer*4+n_simd, i15]] += t2[i15, n_simd]
end n_outer, n_simd
\end{lstlisting}

Listing~\ref{lst:helmholtz_vector_omp} shows the generated C code for the Helmholtz operator vectorized by grouping together four elements. Apart from the previously mentioned changes, we note the following details:
\begin{lstlisting}[language=MyC,
basicstyle=\scriptsize\ttfamily,
float=*ht,
firstnumber=1,
frame=single,
showspaces=false,
showstringspaces=false,
numbers=left,
breaklines=true,
caption=Global assembly code for action of the Helmholtz operator in C vectorized by batching four elements.,
label=lst:helmholtz_vector_omp]
// ... Constant array declarations ... //

void wrap_helmholtz(int const start, int const end, double *__restrict__ dat0, double const *__restrict__ dat1,
                    double const *__restrict__ dat2, int const *__restrict__ map0, int const *__restrict__ map1)
{
  double form_t1[4] __attribute__ ((aligned (64)));
  double t2[6 * 4] __attribute__ ((aligned (64)));
  // ... More temporary array declarations ... //
  for (int n_outer = (start / 4); n_outer <= ((-4 + end) / 4); ++n_outer) {
    for (int i2 = 0; i2 <= 2; ++i2) {
      for (int i3 = 0; i3 <= 1; ++i3) {
        #pragma omp simd
        for (int n_simd = 0; n_simd <= 3; ++n_simd)
          t3[n_simd + 8 * i2 + 4 * i3] = dat1[2 * map1[12 * n_outer + 3 * n_simd + i2] + i3];
      }
    }
    for (int i6 = 0; i6 <= 5; ++i6) {
      #pragma omp simd
      for (int n_simd = 0; n_simd <= 3; ++n_simd)
        t4[n_simd + 4 * i6] = dat2[map0[24 * n_outer + 6 * n_simd + i6]];
    }
    for (int i1 = 0; i1 <= 5; ++i1) {
      #pragma omp simd
      for (int n_simd = 0; n_simd <= 3; ++n_simd)
        t2[n_simd + 4 * i1] = 0.0;
    }
    #pragma omp simd
    for (int n_simd = 0; n_simd <= 3; ++n_simd) {
      form_t11[n_simd] = 0.0;
      form_t1[n_simd] = -1.0 * t3[n_simd];
      // ... More similar instructions ... //
      form_t8[n_simd] = fabs(form_t7[n_simd]);
    }
    for (int form_ip = 0; form_ip <= 5; ++form_ip) {
      // ... More similar loop nests ... //
      for (int form_j = 0; form_j <= 5; ++form_j) {
        #pragma omp simd
        for (int n_simd = 0; n_simd <= 3; ++n_simd)
          t2[n_simd + 4 * form_j] += (form_t25[form_j] * form_t24[n_simd] + form_t13[n_simd + 4 * form_j] +
                                      form_t27[form_j] * form_t26[n_simd]);
      }
    }
    for (int i15 = 0; i15 <= 5; ++i15)
      for (int n_simd = 0; n_simd <= 3; ++n_simd)
        dat0[map0[24 * n_outer + 6 * n_simd + i15]] += t2[n_simd + 4 * i15];
  }
}
\end{lstlisting}
\begin{itemize}
\item The \texttt{n\_simd} loops are pushed to the innermost level. Moreover, this transformation vector-expands temporary arrays such as \texttt{t2}, \texttt{t3}, \texttt{t4} by four, with the expanded dimension labeled as varying the fastest when viewed from (linear) system memory. This ensures their accesses in the \texttt{n\_simd} loops always have unit stride.
\item Loopy provides a mechanism to declare arrays to be aligned to specified memory boundaries (64 bytes in this example).
\item The \texttt{n\_simd} loops are decorated with \texttt{\#pragma omp simd} to inform C compilers to generate SIMD instructions. The exception is the writing back to the global array (lines 43--45), which is sequentialized due to potential race conditions, as different mesh entities could share the same degrees of freedom.
\item The remainder loop which handles the cases where the number of elements is non-divisible by four is omitted here for simplicity.
\item After cross-element vectorization, all local assembly instructions (lines 28--42) are inside \texttt{n\_simd} loops, which always have trip counts of four and are unit stride. All loop-varying array accesses are unit stride in the fastest moving dimension. There are no loop-carried dependencies in \texttt{n\_simd} loops. As a result, the \texttt{n\_simd} loops, and therefore all local assembly instructions, are vectorizable without further consideration of dependencies. This is verified by checking the x86 assembly code and running the program with the Intel Software Development Emulator.
\end{itemize}

\subsection{Vector extensions}%
\label{subsection:vector_extensions}

A more direct way to inform the compiler to emit SIMD instructions without depending on OpenMP is to use \textit{vector extensions}\endnote{https://gcc.gnu.org/onlinedocs/gcc/Vector-Extensions.html}, which support vector data types. These were first introduced in the GNU compiler (GCC), but are also supported in recent versions of the Intel C compiler (ICC) and Clang. Analogous mechanisms exist in various vector-type libraries, e.g.\ VCL~\citep{fog2017vcl}. To evaluate and compare with the directive-based approach from Section~\ref{subsection:openmp_pragma}, we created a further code generation target in Loopy to support vector data types. When inames and corresponding array axes are jointly tagged as vector loops, Loopy generates code to compute on data in vector registers directly, instead of scalar loops over the vector lanes. It is worth noting that the initial intermediate representation of the loop is identical in each case, and that the different specializations are achieved through code transformation.
\begin{lstlisting}[language=MyC,
basicstyle=\scriptsize\ttfamily,
float=*htbp,
firstnumber=1,
frame=single,
showspaces=false,
showstringspaces=false,
numbers=left,
breaklines=true,
caption=Global assembly code for action of the Helmholtz operator in C vectorized by four elements (using vector extensions).,
label=lst:helmholtz_vector_ve]
typedef double double4 __attribute__ ((vector_size (32)));
typedef int int4 __attribute__ ((vector_size (16)));

static double4 const _zeros_double4 __attribute__ ((aligned (64))) = { 0.0 };

// ... Constant array declarations ... //

void wrap_helmholtz(int const start, int const end, double *__restrict__ dat0, double const *__restrict__ dat1,
                    double const *__restrict__ dat2, int const *__restrict__ map0, int const *__restrict__ map1)
{
  double4 form_t1 __attribute__ ((aligned (64)));
  // ... Temporary array declarations ... //

  for (int n_outer = (start / 4); n_outer <= ((-4 + end) / 4); ++n_outer) {
    for (int i2 = 0; i2 <= 2; ++i2) {
      for (int i3 = 0; i3 <= 1; ++i3) {
        #pragma omp simd
        for (int n_simd = 0; n_simd <= 3; ++n_simd)
          t3[n_simd + 8 * i2 + 4 * i3] = dat1[2 * map1[12 * n_outer + 3 * n_simd + i2] + i3];
      }
    }
    for (int i6 = 0; i6 <= 5; ++i6) {
      #pragma omp simd
      for (int n_simd = 0; n_simd <= 3; ++n_simd)
        t4[n_simd + 4 * i6] = dat2[map0[24 * n_outer + 6 * n_simd + i6]];
    }

    for (int i1 = 0; i1 <= 5; ++i1)
      t2[i1] = _zeros_double4;

    form_t11 = _zeros_double4;
    form_t1 = -1.0 * t3[0];

    for (int form_ip = 0; form_ip <= 5; ++form_ip) {
      // ... More similar instructions ... //
      #pragma omp simd
      for (int n_simd = 0; n_simd <= 3; ++n_simd)
        form_t8[n_simd] = fabs(form_t7[n_simd]);
      // ... More similar instructions ... //
      for (int form_j = 0; form_j <= 5; ++form_j)
        t2[form_j] += form_t25[form_j] * form_t24 + form_t13[form_j] + form_t27[form_j] * form_t26;
    }

    for (int i15 = 0; i15 <= 5; ++i15)
      for (int n_simd = 0; n_simd <= 3; ++n_simd)
        dat0[map0[24 * n_outer + 6 * n_simd + i15]] += t2[i15][n_simd];
  }
}
\end{lstlisting}
Listing~\ref{lst:helmholtz_vector_ve} shows the C code generated for the Helmholtz operator vectorized by batching four elements using the vector extension target. Here almost all vectorized (innermost) loops for local assembly are replaced by operations on vector variables. For instructions which do not fit the vector computation model, most noticeably the indirect data gathering (lines 18--19 24--25), or instructions containing built-in mathematics functions which are not supported on vector data types (line 37), Loopy generates scalar loops over vector lanes decorated with \texttt{\#pragma omp simd}. In addition, because vector extensions do not automatically broadcast scalars, any vector instruction with a scalar rvalue is modified by adding the zero vector to the expression, as shown in lines 29 and 31.

Compared to Listing~\ref{lst:helmholtz_vector_omp}, using vector extensions removes most of the innermost loops, and the only remaining OpenMP SIMD directives are due to the limitations of vector extensions as explained previously.

\section{Performance Evaluation}%
\label{sec:evaluation}

We follow the performance evaluation methodology of~\citet{luporini2017algorithm} by measuring the assembly time of a range of operators of increasing complexity and polynomial degrees. Due to the large number of combinations of experimental parameters (operators, meshes, polynomial degrees, vectorization strategies, compilers, hyperthreading), we only report an illustrative portion of the results here, with the entire suite of experiments made available on the interactive online repository CodeOcean~\citep{sun2019cross}.

\subsection{Experimental setup}

We performed experiments on a single node of two Intel systems, based on the Haswell and Skylake microarchitectures, as detailed in Table~\ref{tab:hardware}. Firedrake uses MPI for parallel execution where each MPI process handles the assembly for a subset of the domain. Hybrid MPI-OpenMP parallelization is not supported, and we stress that OpenMP pragmas are only used for SIMD vectorization within a single MPI process. Because we observe that hyperthreading usually improves the performance by 5\% to 10\% for our applications, we set the number of MPI processes to the number of logical cores of the CPU to utilize all available computation resources. Experimental results with hyperthreading turned off are available on CodeOcean. Turbo Boost is switched off to mitigate reproducibility problems that might be caused by dynamic thermal throttling. The batch size, i.e., the number of elements grouped together for vectorization, is chosen to be consistent with the SIMD length. We use three C compilers: GCC 7.3, ICC 18.0 and Clang 5.0. The two vectorization strategies described in Section~\ref{sec:vectorization} are tested on all platforms. We use the listed \textit{Base Frequency} to calculate the peak performance in Table~\ref{tab:hardware}. In reality, modern Intel CPUs dynamically reduce frequencies on heavy workloads with AVX2 and AVX512 instructions, which results in lower achievable performance. Running the optimized LINPACK benchmark binary provided by Intel gives a reasonable indication of achievable peak performance for compute-bound applications.

For the benefit of reproducibility, we have archived the specific versions of Firedrake components used for the experimental evaluation on Zenodo~\citep{zenodo2019firedrake}. An installation of Firedrake with components matching the ones used for evaluation in this paper can be obtained following the instruction at \href{https://www.firedrakeproject.org/download.html}{https://www.firedrakeproject.org/download.html}, with the following command:
\begin{lstlisting}[language=bash,basicstyle=\scriptsize\ttfamily,frame=none]
python3 firedrake-install --doi 10.5281/zenodo.2595487
\end{lstlisting}
The evaluation framework is archived at~\citep{sun2019_2590705}.

\begin{table*}
\small\sf\centering
\caption{Hardware specification for experiments}
\label{tab:hardware}
\begin{tabular}{lcc}
  \toprule
  & \textbf{Haswell Xeon E5-2640 v3} & \textbf{Skylake Xeon Gold 6130}\\
  \midrule
  Base frequency & 2.6 GHz & 2.1 GHz\\
  Physical cores & 8 & 16\\
  SIMD instruction set  & AVX2 & AVX512\\
  \texttt{doubles} per SIMD vector & 4 & 8\\
  Cross-element vectorization batch size &4 & 8\\
  FMA\endnote{Fused multiply-add operations.} units per core & 2 & 2\\
  FMA instruction issue per cycle & 2 & 2\\
  Peak performance (double-precision)\endnote{Calculated as $\textit{base frequency}\times\textit{\#cores}\times\textit{SIMD width}\times 2\ \textit{(for FMA)}\times\textit{\#issue per cycle}$}     & 332.8 GFLOP/s & 1075.2 GFLOP/s\\
  System memory & 4$\times$8 GB DDR4-2133 & 2$\times$32 GB DDR4-2666\\
  LINPACK performance (double-precision)\endnote{Intel LINPACK Benchmark.  \textit{https://software.intel.com/en-us/articles/intel-mkl-benchmarks-suite}}    & 262.5 GFLOP/s & 678.8 GFLOP/s\\
  Memory bandwidth\endnote{STREAM triad benchmark, 2 threads per core.} & 38.5 GB/s & 36.6 GB/s\\
  GCC/Clang arch flag &-march=native &-march=native\\
  ICC SIMD flag &-xcore-avx2 &-xcore-avx512 -qopt-zmm-usage=high\\
  Other compiler flags &-O3 -ffast-math -fopenmp & -O3 -ffast-math -fopenmp \\
  Intel Turbo Boost & OFF & OFF \\
  \bottomrule
\end{tabular}
\end{table*}%

\begin{table*}%
\centering
\caption{Operator characteristics and speed-up summary, using GCC with vector extensions. AI: arithmetic intensity (FLOP/byte). D: trip count of loops over degrees of freedom. Q: trip count of loops over quadrature points. H: speed-up over baseline on Haswell, 16 processes, with vector extensions. S: speed-up over baseline on Skylake, 32 processes, with vector extensions.\label{tab:operators}}
{\footnotesize
\begin{tabular}{c|c|ccccc|ccccc|ccccc|ccccc}
  \toprule
&& \multicolumn{5}{c|}{\texttt{tri}} & \multicolumn{5}{c|}{\texttt{quad}} & \multicolumn{5}{c|}{\texttt{tet}} & \multicolumn{5}{c}{\texttt{hex}} \\
& P & AI & D & Q & H & S & AI & D & Q & H & S & AI & D & Q & H & S & AI & D & Q & H & S \\
\midrule
\multirow{6}{*}{\begin{sideways}\small mass\end{sideways}}
 & 1 & 1.2 & 3 & 3 & 1.0 & 1.0 & 4.7 & 2 & 3 & 1.1 & 1.5 & 2.7 & 4 & 4 & 1.2 & 0.7 & 16.9 & 2 & 3 & 1.8 & 2.8\\
 & 2 & 1.7 & 6 & 6 & 1.3 & 1.0 & 3.9 & 3 & 4 & 0.8 & 1.0 & 5.9 & 10 & 14 & 1.7 & 2.4 & 10.8 & 3 & 4 & 1.1 & 1.5\\
 & 3 & 3.0 & 10 & 12 & 2.0 & 1.3 & 3.9 & 4 & 5 & 0.8 & 1.0 & 8.7 & 20 & 24 & 0.9 & 1.8 & 8.5 & 4 & 5 & 1.8 & 2.5\\
 & 4 & 5.6 & 15 & 25 & 2.4 & 2.6 & 3.9 & 5 & 6 & 2.2 & 1.9 & 39.2 & 35 & 125 & 1.0 & 1.6 & 7.4 & 5 & 6 & 2.1 & 2.8\\
 & 5 & 7.5 & 21 & 36 & 1.1 & 2.0 & 3.9 & 6 & 7 & 2.3 & 1.5 & 55.9 & 56 & 216 & 0.7 & 1.0 & 7.0 & 6 & 7 & 2.0 & 2.7\\
 & 6 & 9.7 & 28 & 49 & 0.8 & 1.6 & 4.1 & 7 & 8 & 2.5 & 1.9 & 81.2 & 84 & 343 & 1.1 & 1.9 & 6.9 & 7 & 8 & 2.2 & 2.7\\
\midrule
\multirow{6}{*}{\begin{sideways}\small helmholtz\end{sideways}}
 & 1 & 1.8 & 3 & 3 & 1.2 & 1.0 & 10.7 & 2 & 3 & 2.0 & 2.9 & 3.9 & 4 & 4 & 1.6 & 1.6 & 45.5 & 2 & 3 & 2.5 & 3.5\\
 & 2 & 5.7 & 6 & 6 & 2.2 & 1.7 & 10.6 & 3 & 4 & 1.0 & 1.3 & 27.3 & 10 & 14 & 2.3 & 5.5 & 34.9 & 3 & 4 & 1.8 & 3.3\\
 & 3 & 9.6 & 10 & 12 & 2.3 & 5.6 & 10.5 & 4 & 5 & 1.3 & 2.1 & 37.5 & 20 & 24 & 1.5 & 3.3 & 27.9 & 4 & 5 & 1.8 & 3.2\\
 & 4 & 17.8 & 15 & 25 & 2.3 & 5.2 & 10.5 & 5 & 6 & 3.2 & 5.7 & 164.1 & 35 & 125 & 1.9 & 2.8 & 24.5 & 5 & 6 & 2.8 & 4.7\\
 & 5 & 23.3 & 21 & 36 & 1.7 & 3.5 & 10.4 & 6 & 7 & 2.8 & 4.8 & 230.1 & 56 & 216 & 1.4 & 1.8 & 23.1 & 6 & 7 & 2.8 & 4.5\\
 & 6 & 29.9 & 28 & 49 & 1.3 & 2.8 & 10.9 & 7 & 8 & 3.0 & 5.0 & 331.4 & 84 & 343 & 1.5 & 4.1 & 22.7 & 7 & 8 & 2.9 & 4.4\\
\midrule
\multirow{6}{*}{\begin{sideways}\small laplacian\end{sideways}}
 & 1 & 0.5 & 3 & 1 & 1.0 & 1.1 & 7.9 & 2 & 3 & 1.7 & 2.7 & 1.9 & 4 & 1 & 1.0 & 1.0 & 37.8 & 2 & 3 & 2.3 & 3.2\\
 & 2 & 2.7 & 6 & 3 & 1.7 & 1.4 & 8.7 & 3 & 4 & 1.0 & 1.2 & 10.4 & 10 & 4 & 2.2 & 3.6 & 27.1 & 3 & 4 & 1.9 & 2.8\\
 & 3 & 4.0 & 10 & 6 & 2.2 & 2.1 & 8.4 & 4 & 5 & 1.5 & 2.0 & 24.0 & 20 & 14 & 2.2 & 3.5 & 21.6 & 4 & 5 & 1.5 & 2.0\\
 & 4 & 6.9 & 15 & 12 & 2.4 & 2.8 & 8.3 & 5 & 6 & 3.1 & 2.9 & 31.5 & 35 & 24 & 2.7 & 3.2 & 19.2 & 5 & 6 & 2.6 & 3.7\\
 & 5 & 12.6 & 21 & 25 & 2.1 & 3.1 & 8.2 & 6 & 7 & 2.9 & 3.9 & 124.3 & 56 & 125 & 2.9 & 2.6 & 18.4 & 6 & 7 & 2.5 & 3.7\\
 & 6 & 16.8 & 28 & 36 & 1.9 & 2.7 & 8.6 & 7 & 8 & 2.8 & 4.0 & 189.3 & 84 & 216 & 2.6 & 2.3 & 18.3 & 7 & 8 & 2.5 & 4.1\\
\midrule
\multirow{6}{*}{\begin{sideways}\small elasticity\end{sideways}}
 & 1 & 0.5 & 3 & 1 & 1.0 & 1.0 & 10.2 & 2 & 3 & 1.9 & 2.8 & 1.9 & 4 & 1 & 1.1 & 1.0 & 48.0 & 2 & 3 & 2.3 & 3.3\\
 & 2 & 3.0 & 6 & 3 & 1.8 & 1.5 & 10.2 & 3 & 4 & 0.9 & 1.3 & 11.6 & 10 & 4 & 2.0 & 6.3 & 31.8 & 3 & 4 & 1.9 & 2.9\\
 & 3 & 4.4 & 10 & 6 & 2.1 & 2.2 & 9.5 & 4 & 5 & 1.5 & 2.0 & 25.6 & 20 & 14 & 2.2 & 3.3 & 24.4 & 4 & 5 & 1.5 & 1.9\\
 & 4 & 7.3 & 15 & 12 & 2.5 & 3.6 & 9.2 & 5 & 6 & 3.0 & 3.9 & 32.7 & 35 & 24 & 2.8 & 3.1 & 21.3 & 5 & 6 & 2.6 & 3.7\\
 & 5 & 13.2 & 21 & 25 & 2.1 & 3.2 & 9.0 & 6 & 7 & 2.8 & 3.9 & 127.5 & 56 & 125 & 2.8 & 2.6 & 20.1 & 6 & 7 & 2.5 & 3.8\\
 & 6 & 17.4 & 28 & 36 & 1.9 & 2.8 & 9.3 & 7 & 8 & 2.8 & 4.3 & 192.6 & 84 & 216 & 2.6 & 2.4 & 19.8 & 7 & 8 & 2.4 & 4.2\\
\midrule
\multirow{6}{*}{\begin{sideways}\small hyperelasticity\end{sideways}}
 & 1 & 0.5 & 3 & 1 & 1.5 & 1.4 & 31.9 & 2 & 4 & 1.2 & 1.9 & 1.7 & 4 & 1 & 1.6 & 1.9 & 183.0 & 2 & 4 & 1.2 & 1.9\\
 & 2 & 9.8 & 6 & 6 & 2.6 & 4.2 & 31.3 & 3 & 6 & 1.9 & 3.0 & 63.2 & 10 & 14 & 3.1 & 5.9 & 137.8 & 3 & 6 & 2.3 & 4.3\\
 & 3 & 26.8 & 10 & 25 & 3.1 & 6.7 & 30.6 & 4 & 8 & 1.4 & 1.6 & 313.8 & 20 & 125 & 3.0 & 6.1 & 118.6 & 4 & 8 & 1.5 & 1.7\\
 & 4 & 40.7 & 15 & 49 & 3.3 & 7.3 & 30.6 & 5 & 10 & 3.3 & 6.3 & 600.0 & 35 & 343 & 2.9 & 4.2 & 110.3 & 5 & 10 & 3.2 & 6.0\\
 & 5 & 56.1 & 21 & 81 & 2.6 & 6.1 & 30.5 & 6 & 12 & 3.4 & 6.6 & 915.2 & 56 & 729 & 2.6 & 2.8 & 108.1 & 6 & 12 & 3.0 & 5.5\\
 & 6 & 74.5 & 28 & 121 & 2.3 & 4.3 & 31.8 & 7 & 14 & 3.3 & 5.9 & 1428.3 & 84 & 1331 & 1.7 & 4.6 & 108.8 & 7 & 14 & 3.0 & 5.7\\
  \bottomrule
\end{tabular}
}
\end{table*}

We measure the execution time of assembling the residual for five operators: the mass matrix (``\texttt{mass}''), the Helmholtz equation (``\texttt{helmholtz}''), the vector Laplacian (``\texttt{laplacian}''), an elastic model (``\texttt{elasticity}''), and a hyperelastic model (``\texttt{hyperelasticity}''). The mathematical description of the operators is detailed in the supplemental material. These operators stem from real-world applications and cover a wide range of complexity: the generated C code for the corresponding global assembly kernels exceeds hundreds of KB for the \texttt{hyperelasticity} operator at high polynomial degree.

We performed experiments on both 2D and 3D domains, with two types of mesh used for each case: triangles (``\texttt{tri}'') and quadrilaterals (``\texttt{quad}'') for 2D problems, tetrahedra (``\texttt{tet}'') and hexahedra (``\texttt{hex}'') for 3D problems. This large variety underscores the broad applicability of our approach. The \textit{arithmetic intensities} and other pertinent characteristics of the operators are listed in Table~\ref{tab:operators}. The memory footprint is calculated assuming perfect caching -- it is thus a lower bound which results in an upper bound estimation for the arithmetic intensity. The triangular and tetrahedral meshes use an affine coordinate transformation (requiring only one Jacobian evaluation per element). The quadrilateral and hexahedral meshes use a bilinear (trilinear) coordinate transformation (requiring Jacobian evaluation at every quadrature point), which usually results in higher arithmetic intensities at low orders. In Firedrake, tensor-product elements~\citep{McRae2014a} benefit from optimizations such as sum factorization to achieve lower asymptotic algorithmic complexity. They are therefore more competitive for higher order methods~\citep{homolya2017exposing}.

We record the maximum execution time of the generated global assembly kernels on all MPI processes. This time does not includes the time in synchronization and MPI data exchange for halo updates. Each experiment is run five times, and the average execution time is reported. Exclusive access to the compute nodes is ensured and threads are pinned to individual logical cores. Startup costs such as code generation time and compilation time are excluded. We use automatic vectorization by GCC without batching, compiled with the optimization flags of Table~\ref{tab:hardware}, as the baseline for comparison. Comparing with our cross-element strategy, the baseline represents the out-of-the-box performance of compiler auto-vectorization for the local element kernel. We note that cross-element vectorization does not alter the algorithm of local assembly except for the vector expansion, as illustrated by Listing~\ref{lst:helmholtz_local} and Listing~\ref{lst:helmholtz_vector_omp}. Consequently, the total number of floating-point operations remains the same. The performance benefit from cross-element vectorization is therefore composable with the operation-reduction optimizations performed by the form compiler to the local assembly kernels.

\subsection{Experimental results and discussion}

\begin{figure*}[htbp!]
\centering
  \includegraphics[width=\textwidth,keepaspectratio]{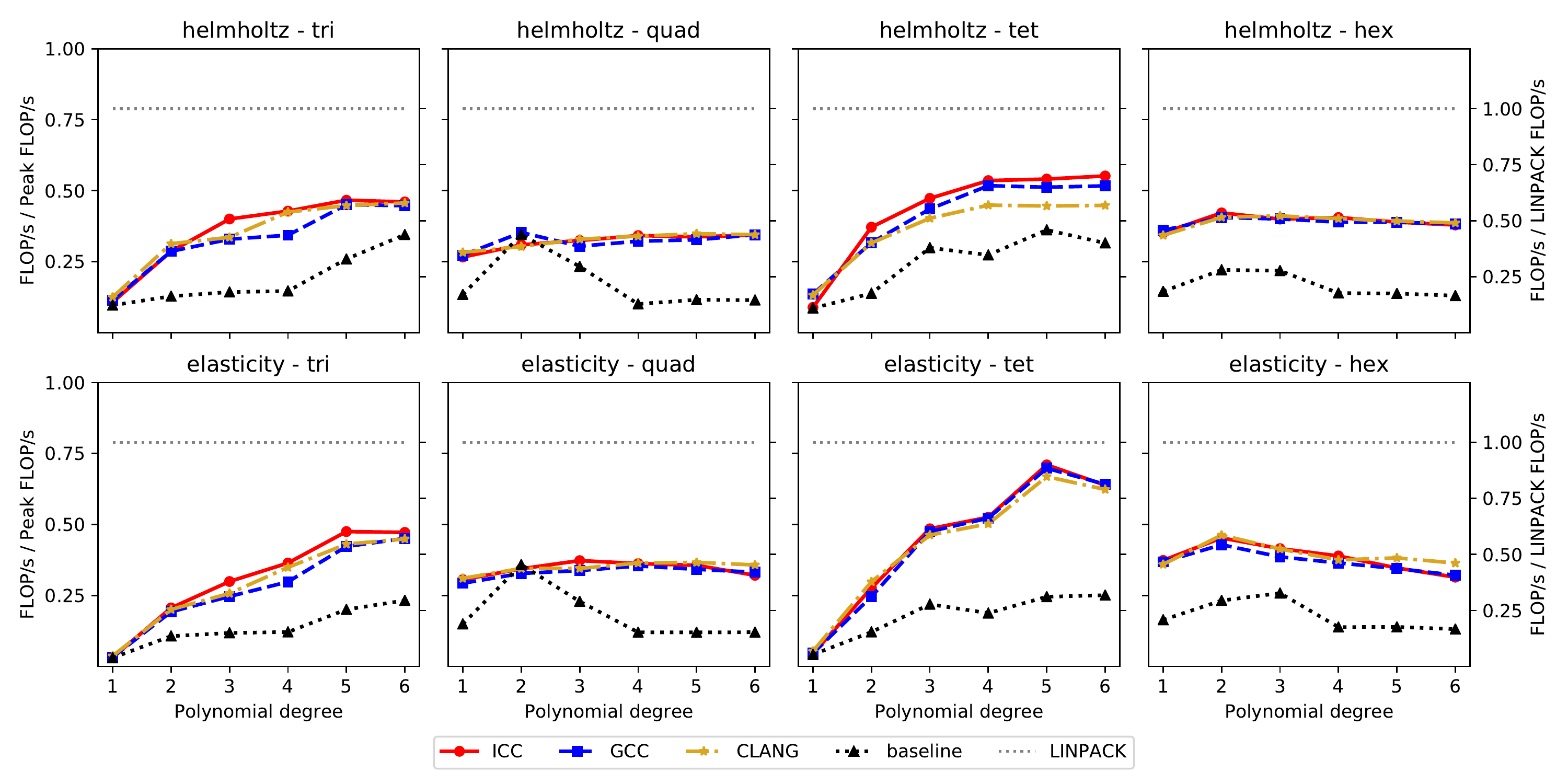}
  \caption{The fraction of peak FLOP/s (as listed in Table~\ref{tab:hardware}) achieved by different compilers for operators \{\texttt{helmholtz, elasticity}\}, on meshes \{\texttt{tri, quad, tet, hex}\} on \textbf{Haswell} using \textbf{vector extensions} with 16 MPI processes. The dotted line indicates the fraction of peak performance achieved by LINPACK benchmark.}%
  \label{fig:haswell_ve}
\end{figure*}

\begin{figure*}[htbp!]
\centering
  \includegraphics[width=\textwidth,keepaspectratio]{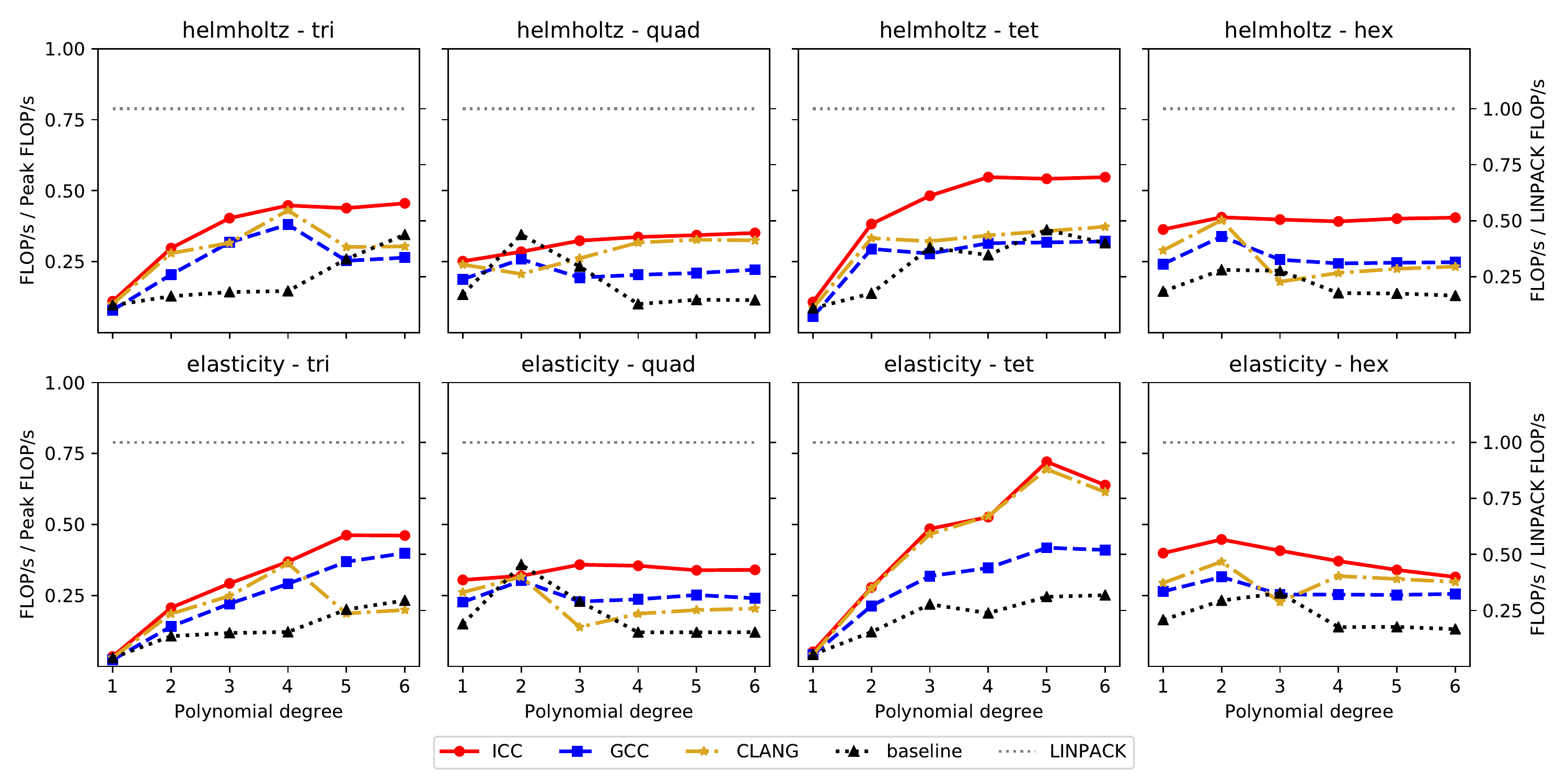}
  \caption{The fraction of peak FLOP/s (as listed in Table~\ref{tab:hardware}) achieved by different compilers for operators \{\texttt{helmholtz, elasticity}\}, on meshes \{\texttt{tri, quad, tet, hex}\} on \textbf{Haswell} using \textbf{OpenMP SIMD directives} with 16 MPI processes. The dotted line indicates the fraction of peak performance achieved by LINPACK benchmark.}%
  \label{fig:haswell_omp}
\end{figure*}

\begin{figure*}[htbp!]
\centering
  \includegraphics[width=\textwidth,keepaspectratio]{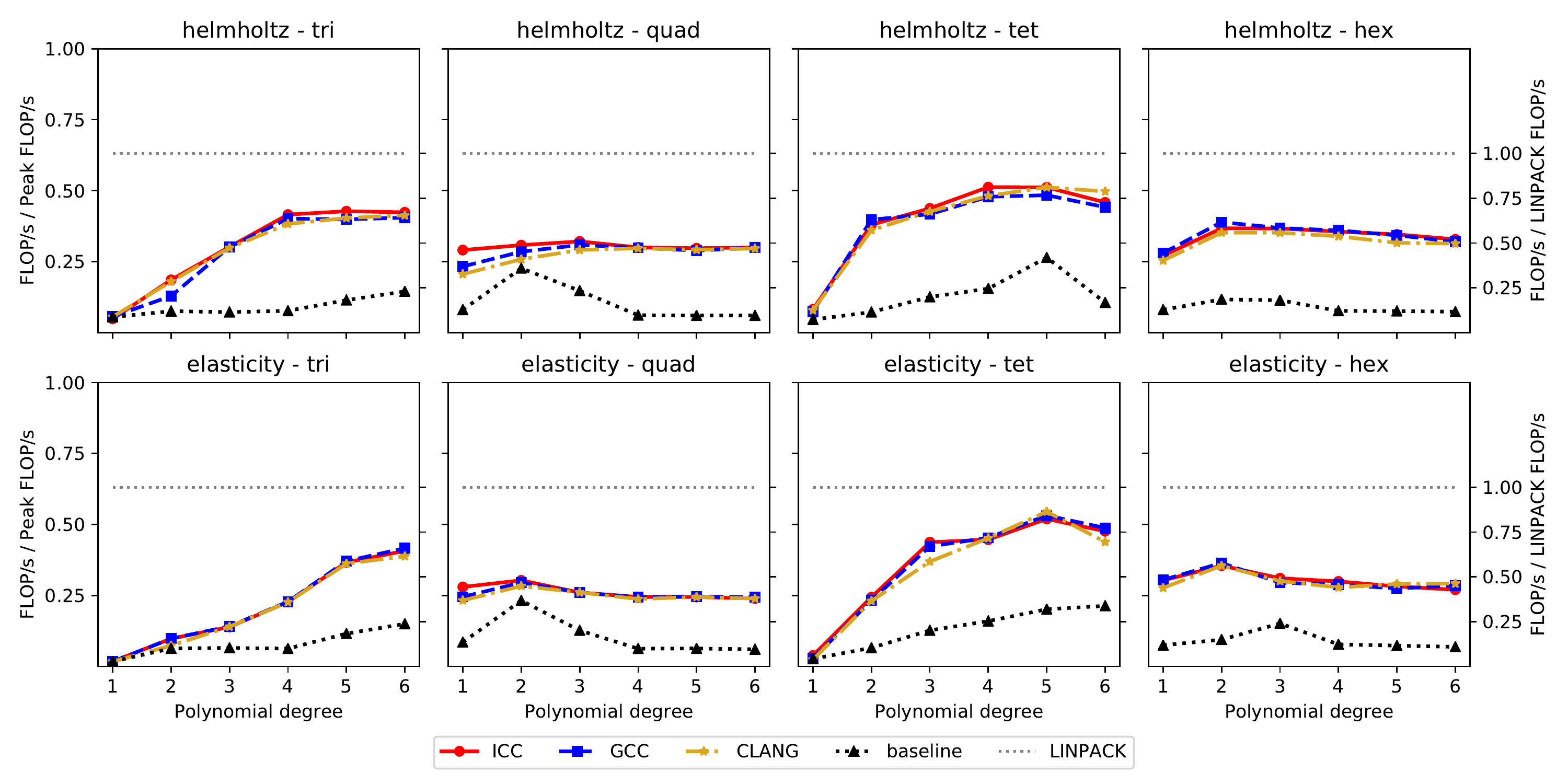}
  \caption{The fraction of peak FLOP/s (as listed in Table~\ref{tab:hardware}) achieved by different compilers for operators \{\texttt{helmholtz, elasticity}\}, on meshes \{\texttt{tri, quad, tet, hex}\} on \textbf{Skylake} using \textbf{vector extensions} with 32 MPI processes. The dotted line indicates the fraction of peak performance achieved by the LINPACK benchmark.}%
  \label{fig:skylake_ve}
\end{figure*}

\begin{figure*}[htbp!]
\centering
  \includegraphics[width=\textwidth,keepaspectratio]{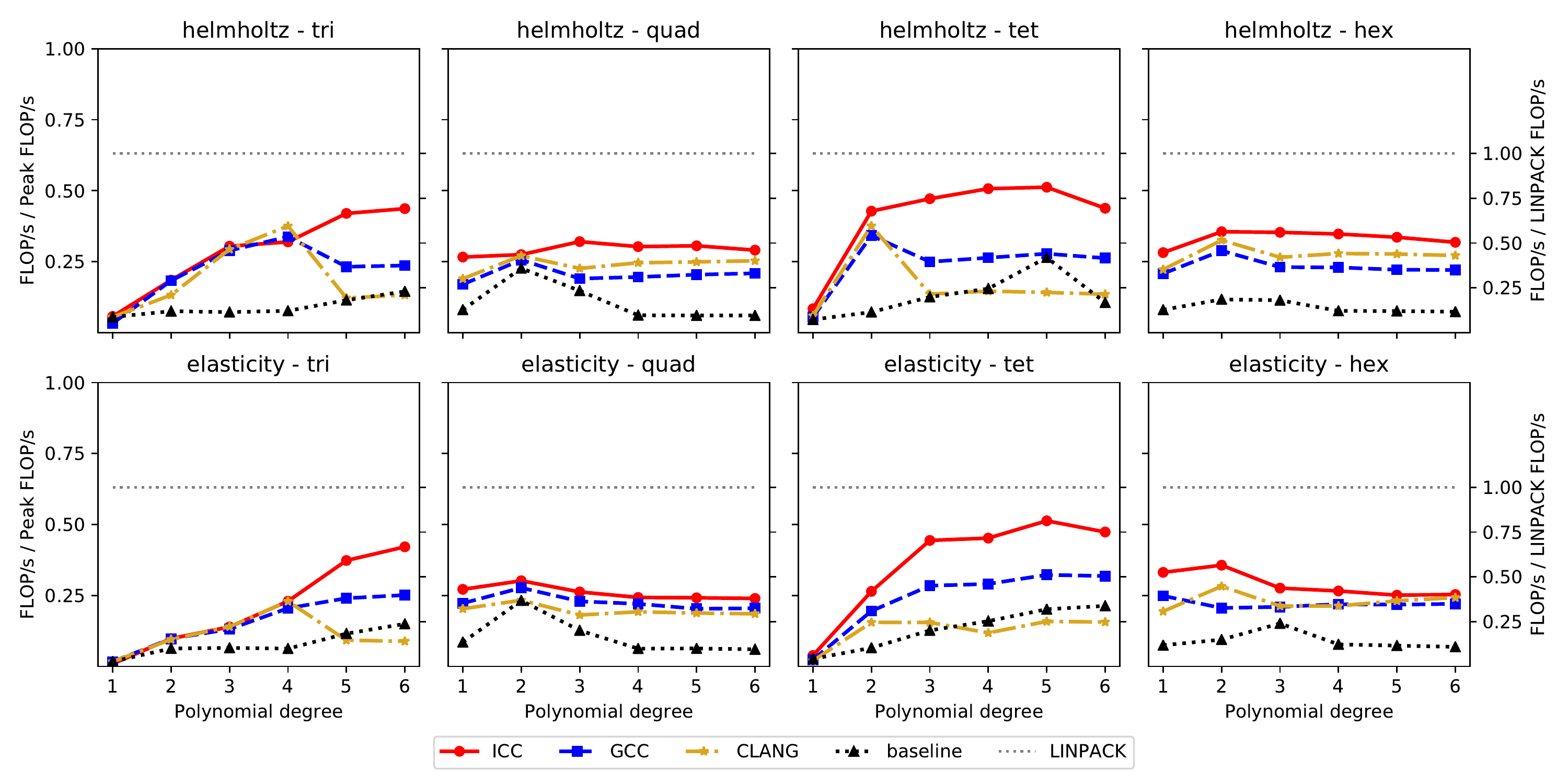}
  \caption{The fraction of peak FLOP/s (as listed in Table~\ref{tab:hardware}) achieved by different compilers for operators \{\texttt{helmholtz, elasticity}\}, on meshes \{\texttt{tri, quad, tet, hex}\} on \textbf{Skylake} using \textbf{OpenMP SIMD directives} with 32 MPI processes. The dotted line indicates the fraction of peak performance achieved by the LINPACK benchmark.}%
  \label{fig:skylake_omp}
\end{figure*}

\begin{figure*}[htbp!]
\centering
  \includegraphics[width=\textwidth,keepaspectratio]{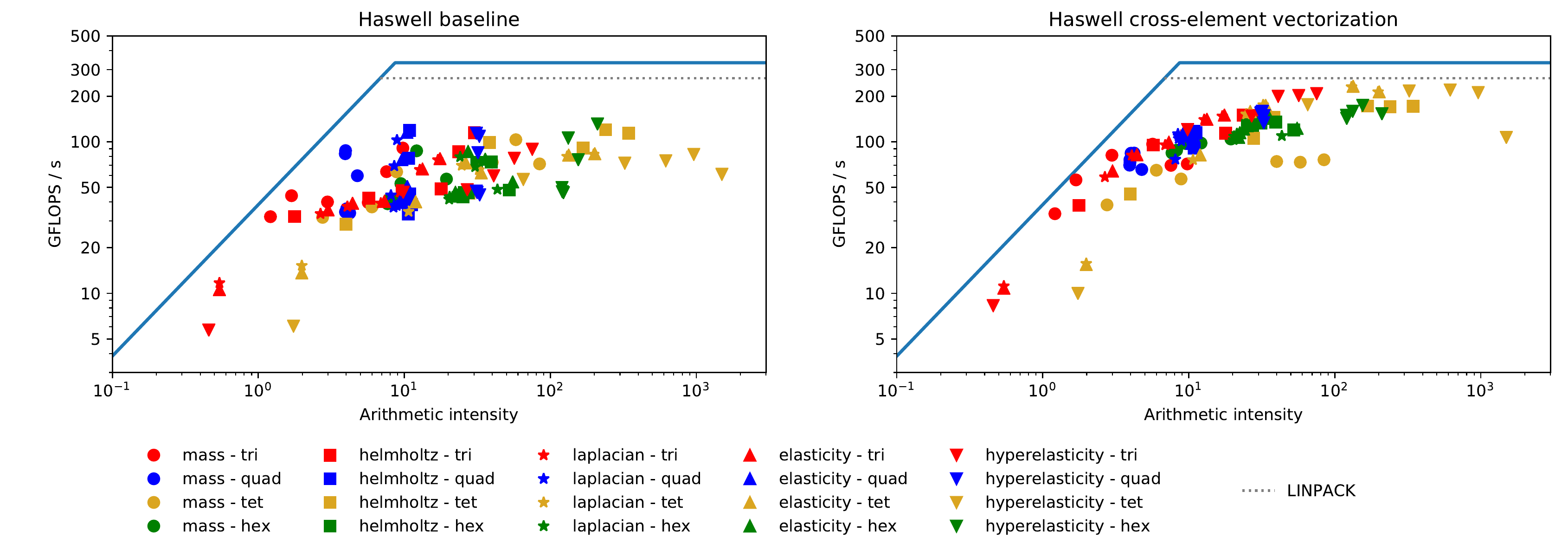}
  \caption{Roofline model of operators for baseline and cross-element vectorization using GCC on \textbf{Haswell}. The dotted lines indicate the performance of the LINPACK benchmark.}%
  \label{fig:roofline_haswell}
\end{figure*}

\begin{figure*}[ht!]
  \includegraphics[width=\textwidth,keepaspectratio]{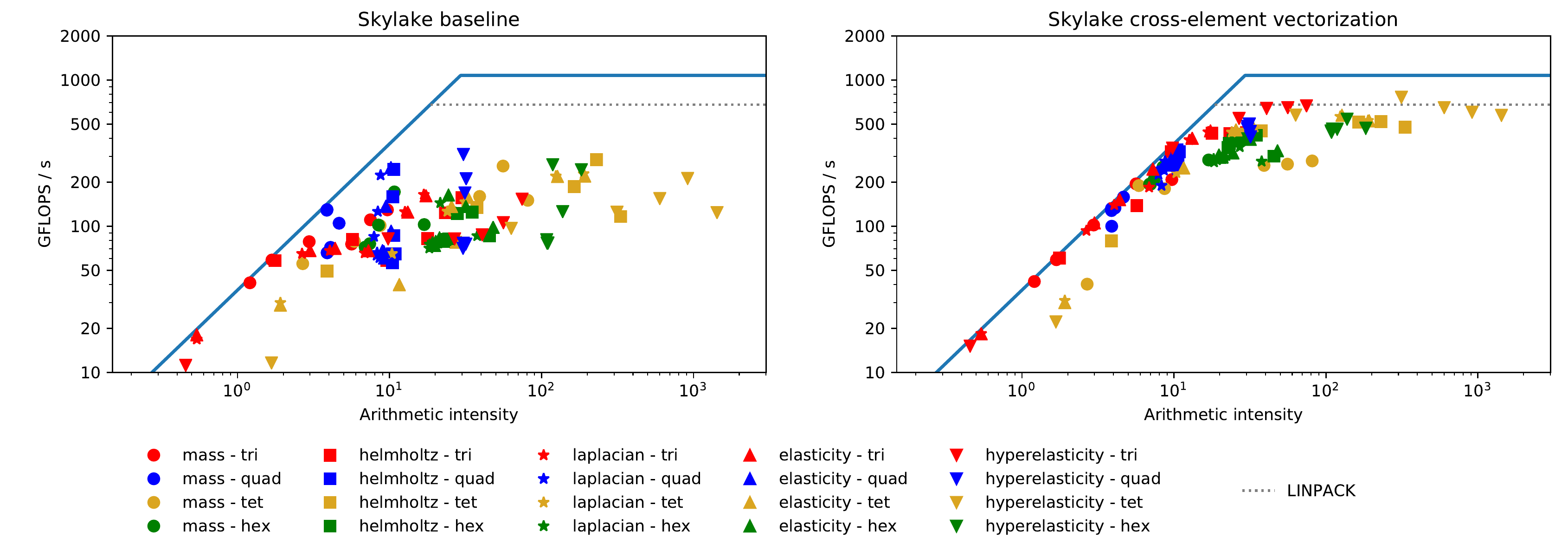}
  \caption{Roofline model of operators for baseline and cross-element vectorization using GCC on \textbf{Skylake}. The dotted lines indicate the performance of the LINPACK benchmark.}%
  \label{fig:roofline_skylake}
\end{figure*}

Figures~\ref{fig:haswell_ve} to~\ref{fig:skylake_omp} show the performance of the \texttt{helmholtz} and \texttt{elasticity} operators on Haswell and Skylake, vectorized with OpenMP SIMD directives (Section~\ref{subsection:openmp_pragma}), and with vector extensions (Section~\ref{subsection:vector_extensions}). We indicate the fraction of peak performance achieved on the left axis, and the fraction of the LINPACK benchmark performance on the right axis. Figures~\ref{fig:roofline_haswell} and~\ref{fig:roofline_skylake} compare roofline models~\citep{williams2009roofline} of the baseline and our approach using vector extensions and the GCC compiler on Haswell and Skylake. The speed-up achieved is also summarized in Table~\ref{tab:operators}.

\subsubsection{Compiler comparison and vector extensions}

When vectorizing with OpenMP SIMD directives, ICC gives the best performance for almost all test cases, followed by Clang, while GCC is significantly less competitive. The performance disparity is more pronounced on Skylake than on Haswell. However, when using vector extensions, Clang and GCC improve significantly and are able to match the performance of ICC on both Haswell and Skylake, whereas ICC performs similarly with both OpenMP SIMD directives and vector extensions.

We use the Intel Software Development Emulator\endnote{https://software.intel.com/en-us/articles/intel-software-development-emulator} to count the number of instructions executed at runtime for code generated by different compilers. The data indicate that although floating-point operations are fully vectorized by all compilers, GCC and Clang generate more load and store instructions between vector registers and memory when using OpenMP SIMD directives for vectorization. One possible reason is that GCC and Clang choose to allocate short arrays to the stack rather than the vector registers directly, causing more load on the memory subsystem.

In light of these results, we conclude that vectorization with vector extensions allows greater performance portability on different compilers and CPUs for our application. It is, therefore, our preferred strategy for implementing cross-element vectorization, and is the default option for the rest of our analysis.

\subsubsection{Vectorization speed-up}

Almost across the board, significant speed-up is achieved on the test cases under consideration. Slowdown occurs in two situations. On low polynomial degrees, the kernels tend to have low arithmetic intensity so that the increase in available floating point throughput through cross-element vectorization cannot compensate for the increase in the size of the working set of data. On simple operators such as \texttt{mass} on \texttt{tri} and \texttt{tetra}, the kernels have simple loop structures and the compilers can sometimes successfully apply other optimizations such as unrolling and loop interchange to achieve vectorization without batching elements in the outer loop. The pattern of speed-up is consistent across Haswell and Skylake. Higher speed-up is generally achieved on more complicated operators (e.g. \texttt{hyperelasticity}), and on tensor-product elements (\texttt{quad} and \texttt{hex}), which generally correspond to more complicated loop structure and higher arithmetic intensity due to the Jacobian recomputation at each quadrature point.

\subsubsection{Achieved fraction of peak performance}

We observe that the fraction of peak performance varies smoothly with polynomial degrees for cross-element vectorization in all test cases. This fulfils an important design requirement for Firedrake: small changes in problem setup by the users should not create unexpected performance degradation. This is also shown in Figures~\ref{fig:roofline_haswell} and~\ref{fig:roofline_skylake} where the results are more clustered on the roofline plots after cross-element vectorization. The baseline shows performance inconsistency, especially for low polynomial degrees. For instance, for the \texttt{helmholtz} operator with degree 3 on \texttt{quad}, the quadrature loops and the basis function loops all have trip counts of 4, which fits the vector length on Haswell and results in better performance.

On simplicial meshes (\texttt{tri} and \texttt{tetra}), higher order discretization leads to kernels with very high arithmetic intensity because of the quadratic and cubic increases in the number of basis functions, and thus the loop trip counts. This is due to the current limitation that simplicial elements in Firedrake are not sum factorized. In these test cases, we observe that the baseline approaches cross-element vectorization for sufficiently high polynomial degrees. This is not a serious concern for our optimization approach because the break-even degrees are very high except for simple operators such as \texttt{mass}, and ultimately tensor-product elements are more competitive for higher order methods in terms of algorithmic complexity. 

We also observe that there exist a small number of test cases where the achieved peak performance is marginally higher than the LINPACK benchmark on Skylake, as shown in Figure~\ref{fig:roofline_skylake}. One possible reason for this observation is thermal throttling since our test cases typically run for a shorter period of time than LINPACK\@. We also note that these test cases correspond to high order \texttt{hyperelasticity} operators on \texttt{tet} meshes, which, as noted previously, are not practically important use cases since using tensor-product elements requires many fewer floating-point operations at the same polynomial order.

\subsubsection{Tensor-product elements}

We observe higher and more consistent speed-up for tensor-product elements (\texttt{quad} and \texttt{hex}) on both Haswell and Skylake. This is because, on these meshes, more computation is moved out of the innermost loop due to sum factorization, resulting in more challenging loop nests for the baseline strategy which attempts to vectorize within the element kernel. The same applies to the evaluation of the Jacobian of coordinate transformation, which is a nested loop over quadrature points after sum factorization for tensor-product elements.

The base elements of \texttt{quad} and \texttt{hex} are interval elements in 1D, thus the extents of loops over degrees of freedom increase only linearly with respect to polynomial degrees, as shown in Table~\ref{tab:operators}. As a result, the baseline performance does not improve as quickly for higher polynomial degrees on \texttt{quad} and \texttt{hex} compared with \texttt{tri} and \texttt{tet}, resulting in stable speed-up for cross-element vectorization observed on tensor-product elements.

\section{Conclusion and future work}%
\label{sec:conclusion}

We have presented a portable, general-purpose solution for delivering stable vectorization performance on modern CPUs for matrix-free finite element assembly for a very broad class of finite element operators on a large range of elements and polynomial degrees. We described the implementation of cross-element vectorization in Firedrake which is transparent to the end users. Although the technique of cross-element vectorization is conceptually simple and has been applied in hand-written kernels before, our implementation based on code generation is automatic, robust and composable with other optimization passes.

We have presented extensive experimental results on two recent Xeon processors that are commonly used in HPC applications, and compared the vectorization performance of three popular C compilers. We showed that by generating appropriate vectorizable code, and using compiler-based vector extensions, we can obtain portably high performance across all three compilers.

The write-back to global data structure is not vectorized in our approach due to possible race conditions. The newly introduced Conflict Detection instructions in the Intel AVX512 instruction set could potentially mitigate this limitation~\citep[Section 2.3]{zhang2016guide}. This could be achieved by informing Loopy to use the relevant intrinsics when generating code for loops with specific tags.

We have focused on the matrix-free finite element method because it is compute-intensive and more likely to benefit from vectorization. However, our methods and implementation also support matrix assembly. Firedrake relies on PETSc~\citep{balay2017petsc} to handle distributed sparse matrices, and PETSc requires certain data layouts for the input array when updating the global matrices. When several elements are batched together for cross-element vectorization, we need to generate code to explicitly unpack/transpose the local assembly results into individual arrays before calling PETSc functions to update the global sparse matrices for each element. Future improvement could include eliminating this overhead, possibly by extending the PETSc API\@.

The newly introduced abstraction layer, together with Loopy integration in the code generation and optimization pipeline, opens up multiple possibilities for future research in Firedrake. These include code generation with intrinsics instructions, loop tiling, and GPU acceleration, all of which are already supported in Loopy.

\begin{acks}
The authors would like to thank Tobias Grosser, Richard Veras, J.~Ramanujam and P.~Sadayappan for their valuable insights during our discussions which started at Dagstuhl Seminar 18111 on Loop Optimization. The authors are grateful to James Cownie and Andrew Mollinson at Intel Corp.\ as well as Koki Sagiyama at Imperial College London for providing access to the Skylake platform.
\end{acks}

\begin{dci}
The authors declared no potential conflicts of interest with respect to the research, authorship, and/or publication of this article.
\end{dci}

\begin{funding}
This work was supported by the Engineering and Physical Sciences Research Council [grant numbers EP/L016796/1, EP/R029423/1], and the Natural Environment Research Council [grant number NE/K008951/1]. It was further funded by the US Navy Office of Naval Research under grant number N00014-14-1-0117 and the US National Science Foundation under grant number CCF-1524433. AK gratefully acknowledges a hardware gift from Nvidia Corporation.

\end{funding}

\begin{sm}%
\label{sm:operators}
Here we describe the operators used as the test cases for experimental evaluation. They are defined as bilinear forms, and we take their \texttt{action} in UFL to obtain the corresponding linear forms.

\begin{description}
\item[\texttt{mass}] Here $u$ and $v$ are scalar-valued trial and test functions.
\begin{equation}
a = \int uv\ \mathrm{d}x
\end{equation}

\item[\texttt{helmholtz}] Here $u$ and $v$ are scalar-valued trial and test functions.
\begin{equation}
a = \int (\nabla u \cdot \nabla v\ + uv)\ \mathrm{d}x
\end{equation}

\item[\texttt{laplacian}] Here $\mathbf{u}$ and $\mathbf{v}$ are vector-valued trial and test functions.
\begin{equation}
a = \int (\nabla \mathbf{u} : \nabla \mathbf{v})\ \mathrm{d}x
\end{equation}

\item[\texttt{elasticity}] The linear elasticity model solves for a displacement vector field. Here $\mathbf{u}$ and $\mathbf{v}$ are vector-valued trial and test functions, $\epsilon$ is the symmetric strain rate tensor. The bilinear form is defined as:
\begin{equation}
\begin{aligned}
\epsilon(\mathbf{u}) &= \frac{1}{2}\big[ \nabla\mathbf{u} + (\nabla\mathbf{u})^T \big] \\
a &=\int \epsilon(\mathbf{u}):\epsilon(\mathbf{v}) \ \mathrm{d}x
\end{aligned}
\end{equation}

\item[\texttt{hyperelasticity}] In this simple hyperelastic model, we define the strain energy function $\Psi$ over vector field $\mathbf{u}$:
\begin{equation}
\begin{aligned}
\mathbf{F} &= \mathbf{I} + \nabla \mathbf{u}\\
\mathbf{C} &= \mathbf{F}^T\mathbf{F}\\
\mathbf{E} &=(\mathbf{C} - \mathbf{I})/2,\\
\Psi &= \frac{\lambda}{2}\big[\textsf{tr}(\mathbf{E})\big]^2+\mu\textsf{tr}(\mathbf{E}^2)
\end{aligned}
\end{equation}
where $\mathbf{I}$ is the identity matrix, $\lambda$ and $\mu$ are the Lam\'{e} parameters of the material, $\mathbf{F}$ is the deformation gradient, $\mathbf{C}$ is the right Cauchy-Green tensor, $\mathbf{E}$ is the Euler-Lagrange strain tensor. We define the Piola-Kirchhoff stress tensors as:
\begin{equation}
\begin{aligned}
\mathbf{S} &= \frac{\partial\Psi}{\partial\mathbf{E}}\\
\mathbf{P} &= \mathbf{F}\mathbf{S}
\end{aligned}
\end{equation}
Finally, we arrive at the residual form of this nonlinear problem:
\begin{equation}
r=\int \mathbf{P}:\nabla \mathbf{v} - \mathbf{b}\cdot\mathbf{v} \ \mathrm{d}x
\end{equation}
where $\mathbf{b}$ is the external forcing. To solve this nonlinear problem, we need to linearize the residual form at an approximate solution $\mathbf{u}$, this gives us the bilinear form $a$:
\begin{equation}
a=\lim_{\epsilon \to 0}\frac{r(\mathbf{u}+\epsilon\delta\mathbf{u})-r(\mathbf{u})}{\epsilon},
\end{equation}
where the trial function is $\delta\mathbf{u}$, the test function is $\mathbf{v}$, and $\mathbf{u}$ is a coefficient of the operator. We use the automatic differentiation of UFL to compute the operator symbolically.

\end{description}

\end{sm}

\theendnotes

\bibliographystyle{SageH}

%\bibliography{vector-ijhpca}

\end{document}